\documentclass[aps,twocolumn,amsmath,amssymb,superscriptaddress,prb]{revtex4-1}
\usepackage{graphicx}% Include figure files
\usepackage{dcolumn}% Align table columns on decimal point
\usepackage{subfigure}
\usepackage{bm}% bold math

\begin{document}

%\preprint{APS/123-QED}

\title{Electronic Structure and Lattice Dynamics in the FeSb$_{3}$ Skutterudite from Density Functional Theory}

\author{Mikael R{\aa}sander}\email{mikra@kth.se}
\affiliation{%
Department of Materials and Nanophysics, KTH Royal Institute of Technology, Electrum 229, SE-164 40 Kista, Sweden
}%
%\author{Long Bui Doc}
%\affiliation{%
%Department of Materials and Nanophysics, KTH Royal Institute of Technology, Electrum 229, SE-164 40 Kista, Sweden
%}%
%\affiliation{%
%Department of Mechanical Engineering, University of Malaya, 506 03 Kuala Lumpur, Malaysia
%}%
\author{Lars Bergqvist}
\affiliation{%
Department of Materials and Nanophysics, KTH Royal Institute of Technology, Electrum 229, SE-164 40 Kista, Sweden
}%
\affiliation{%
SeRC (Swedish e-Science Research Center), KTH, SE-100 44 Stockholm, Sweden}%
\author{Anna Delin}
\affiliation{%
Department of Materials and Nanophysics, KTH Royal Institute of Technology, Electrum 229, SE-164 40 Kista, Sweden
}%
\affiliation{%
SeRC (Swedish e-Science Research Center), KTH, SE-100 44 Stockholm, Sweden}%
\affiliation{%
Department of Physics and Astronomy, Uppsala University, Box 516, SE-751 20 Uppsala, Sweden
}%
\date{\today}% It is always \today, today,

\begin{abstract}
We have performed density functional calculations of the electronic structure and lattice dynamics of the binary skutterudite FeSb$_{3}$. We find that the ground state of FeSb$_{3}$ is a near semi-metallic ferromagnet with $T_{c}=175$~K. Furthermore, we find that FeSb$_{3}$ is softer than CoSb$_{3}$ based on an analysis of the relation of the elastic constants and the shape of the phonon density of states in the two systems, which is in agreement with experimental observation. Based on these observations we find it plausible that FeSb$_{3}$ will have a lower thermal conductivity than CoSb$_{3}$. Additionally, our calculations indicate that FeSb$_{3}$ may be stable towards decomposition into FeSb$_{2}$ and Sb. Furthermore, for ferromagnetic FeSb$_{3}$ we obtain real-valued phonon frequencies and also a $c_{44}$ greater than zero, indicating that the system is mechanically as well as dynamically stable.
\end{abstract}

%\pacs{}% PACS, the Physics and Astronomy
\maketitle
\section{introduction}\label{sec:intro}
The skutterudites constitute an interesting class of materials for applications as thermoelectric energy converters,\cite{Snyder2008} since they possess the necessary electronic properties, notably a large Seebeck coefficient, and a low thermal conductivity. The latter is largely due to filler atoms that occupy large voids in the crystal structure.\cite{Feldman,Feldman2,Feldman3,Hermann,Kendziora} The exact mechanism of the filler atoms to lower the thermal conductivity is however debated.\cite{Koza} The skutterudites can generally be presented as R$_{\rm y}$M$_{4}$X$_{12}$, where M is a transition metal, such as Co, Ir, Rh. X is a pnictogen, e.g. P, As and Sb, and R is the filler atom, typically a rare earth element, such as La or Ce. The skutterudites offer a rich ability for engineering in order to optimize their properties by alloying on the metal or pnictogen lattices, as well as by careful selection of the filler element.
\par
CoSb$_{3}$ is an archetypical skutterudite system. It has excellent electronic properties and especially it has a high Seebeck coefficient.\cite{Smith,Wee,Hammerscmidt} Unfortunately, the thermal conductivity of CoSb$_{3}$ is too large for achieving a high thermoelectric efficiency. The ability to lower the thermal conductivity in CoSb$_{3}$ by using filler atoms is limited\cite{Shi} since the filling factor, y, is rather small and focus on filled skutterudites has been directed towards other skutterudites, such as Fe containing R$_{\rm y}$Fe$_{\rm x}$Co$_{1-{\rm x}}$Sb$_{12}$ since the filling factor is larger for such systems and for R$_{\rm y}$Fe$_{4}$Sb$_{12}$ the filling factor reaches unity.\cite{Feldman,Feldman2,Feldman3,Hermann,Kendziora} Filled R$_{\rm y}$Fe$_{4}$Sb$_{12}$ systems have been studied for quite some time. However, the binary FeSb$_{3}$ system has not received much attention even though an understanding of the dynamics of the filler and its impact on the electronic structure and thermal conductivity cannot be complete without an understanding of the properties of the host framework. According to the phase diagram\cite{phasediagram} FeSb$_{3}$ is metastable in comparison to FeSb$_{2}$, which also possess interesting thermoelectric properties, and Sb. Recently, however, micrometer thick films of FeSb$_{3}$ has been synthesized by nanoalloying of Fe and Sb precursors at $T\sim400$~K\cite{Hornbostel,Mochel} and its physical properties investigated experimentally.\cite{Mochel} A significant difference compared to CoSb$_{3}$ is that FeSb$_{3}$ is softer and a softening of low-energy phonon modes would likely have a favorable influence towards a lower thermal conductivity of FeSb$_{3}$ compared to CoSb$_{3}$.
\par
Since there are no theoretical studies focusing on FeSb$_{3}$ we investigate the electronic structure, lattice dynamics and possible dynamic stability of FeSb$_{3}$ using methods based on density functional theory. We compare our results with calculations  for the well-known CoSb$_{3}$ system in order to elucidate the differences between these two superficially rather similar compounds. In addition, we will analyze how filling the voids in FeSb$_{3}$ with La affects the lattice dynamics.
%Thus far no theoretical studies have focused on FeSb$_{3}$. 
%There are studies of the lattice dynamics of LaFe$_{4}$Sb$_{12}$ and CeFe$_{4}$Sb$_{12}$ in which lattice dynamical properties of FeSb$_{3}$ are presented.\cite{Feldman} However, in these studies the lattice dynamics is derived from force constant models and the force constants for FeSb$_{3}$ are either taken from data of LaFe$_{4}$Sb$_{12}$, where the FeSb$_{3}$ lattice dynamics were obtained by removal of La, or from data on CoSb$_{4}$, where FeSb$_{3}$ dynamics were obtained by changing the mass~of~Co.
%\par
%Here 
\par
The paper is outlined as follows: In Section~\ref{sec:method} we will present the details of our calculations and in Section~\ref{sec:results} we will present our results. Finally in Section~\ref{sec:conclusions} we will summarize our findings and present our conclusions. 
\begin{figure}[t]
\includegraphics[width=8cm]{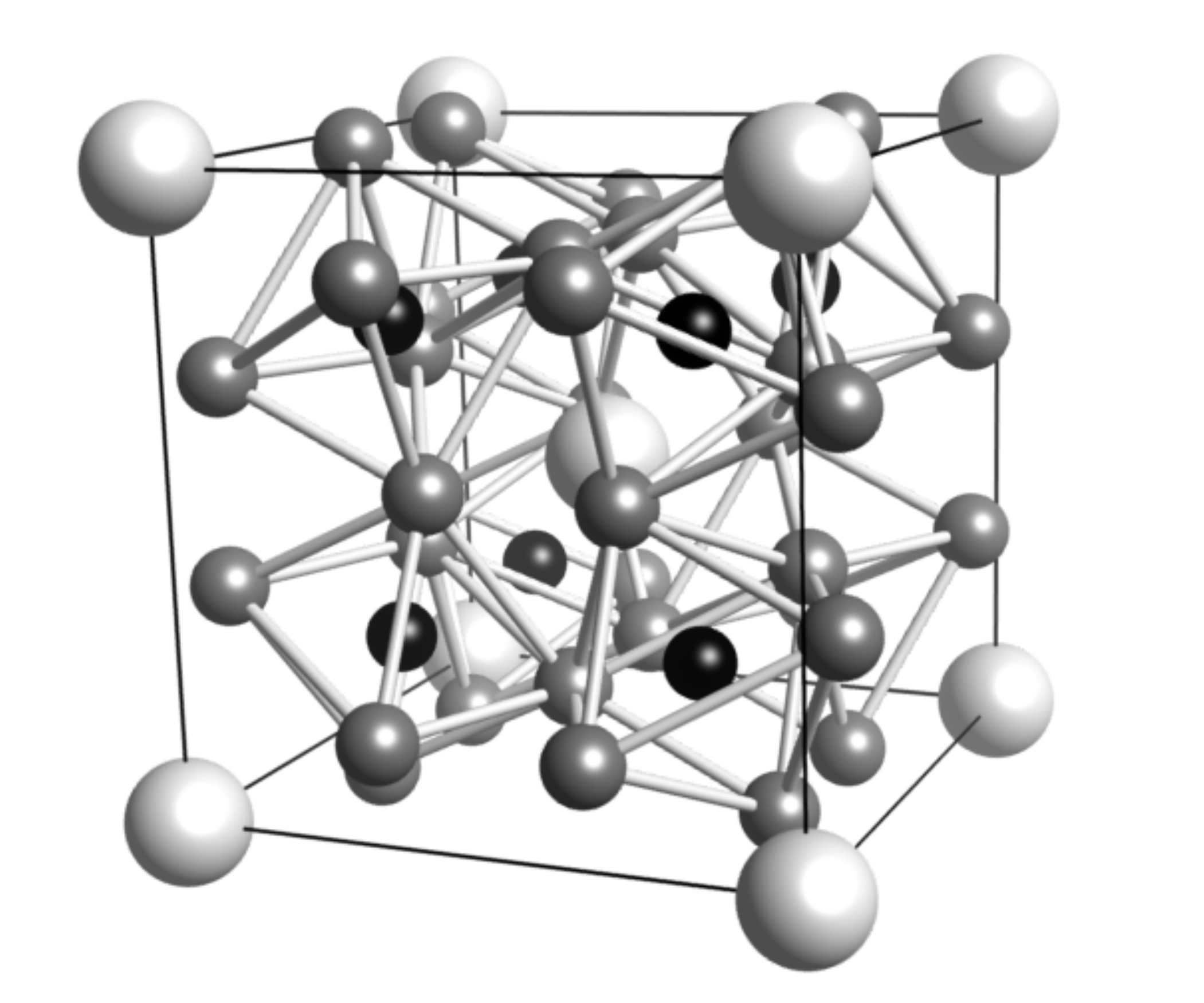}
\caption{\label{fig:structure} Illustration of the crystal structure of the skutterudite structure. Fe (black spheres) is residing inside canted octahedral cages of Sb (grey spheres). Filler atoms are presented by large white spheres. In this case the filler is La.}
\end{figure}

\section{computational details}\label{sec:method}
The binary skutterudite structure has a unit cell containing four formula units with body centered cubic lattice vectors and belongs to the space group Im-3 (No. 204), where metal atoms and pnictogen atoms occupy the 8$c$ and 24$g$ positions respectively. The Sb atoms occupy the general position (0,$y$,$z$) and these values along with our calculated lattice constants are shown in Table~\ref{tab:fesb3}. The skutterudite framework, i.e. MX$_{3}$, contains large voids at the 2$a$ positions of the lattice. Filling these voids by other atoms do not change the symmetry of the crystal. In Fig.~\ref{fig:structure} we show the conventional unit cell of the skutterudite structure (8 formula units) with the voids, at (0,0,0) and ($\frac{1}{2}$,$\frac{1}{2}$,$\frac{1}{2}$), filled with a rare-earth element. In this study the rare-earth element is La, and calculations have been performed for FeSb$_{3}$ and CoSb$_{3}$ binary skutterudite and the filled LaFe$_{4}$Sb$_{12}$.
\par
Density functional calculations have been performed using the projector augmented wave (PAW) method\cite{Blochl} as it is implemented in the Vienna {\it ab-initio} simulation package (VASP)\cite{KresseandFurth,KresseandJoubert}. The generalized gradient approximation due to Perdew, Burke and Ernzerhof (PBE)\cite{PBE} has been used for the exchange-correlation energy functional. Relaxation of the ionic position as well as the volume of the systems have been performed until the forces on individual atoms were smaller than 0.1~meV/\AA. A k-point mesh of $6\times6\times6$\cite{MonkhorstandPack} was found to be accurate enough for obtaining converged total energies and structural parameters. The plane wave energy cut-off was set to 600~eV. Spin-polarization was considered for all systems and it will be made clear when a non spin-polarized (NSP), ferromagnetic (FM) or anti-ferromagnetic (AFM) solution is discussed. Here, we focus on the electronic structure and lattice dynamics of the FeSb$_{3}$ system in relation to the more studied CoSb$_{3}$ and LaFe$_{4}$Sb$_{12}$ systems and a more elaborate study of the magnetic properties of FeSb$_{3}$ will be published elsewhere in conjunction with Co$_{1-x}$Fe$_{x}$Sb$_{3}$ alloyed systems.\cite{magneticFeSb3} However, here we have accurately calculated the Curie temperature (T$_c$) of FeSb$_{3}$ by calculating the exchange parameters within a Heisenberg model using the SPR-KKR package,\cite{SPRKKR} where the crystal geometry was taken from our PAW calculations, and subsequent Monte Carlo simulations using the UppASD package.\cite{UppASD} We have also checked the effects of spin-orbit coupling on the electronic structure and lattice dynamics in FeSb$_{3}$ and CoSb$_{3}$ and found that the effects are rather minor.
\par
The lattice dynamics have been calculated within the harmonic approximation at $T=0$~K by means of the small displacement method, as it is implemented in the Phonopy code.\cite{phonopy} The results shown here are obtained for a 3$\times$3$\times$3 multiplication of the primitive skutterudite unit cell which is considered large enough to yield well converged lattice dynamical properties. The size of the displacements were~0.01~\AA. For the supercells a 2$\times$2$\times$2 k-points mesh has been used for the electronic structure calculations from which the forces acting on the atoms have been evaluated. In the calculations of the phonon density of states (PDOS) a 20$\times$20$\times$20 q-points mesh has been utilized.

\begin{table}[t]
\caption{\label{tab:fesb3} Comparison of the evaluated lattice constants and crystallographic $y$ and $z$ values for the Sb atoms in FeSb$_{3}$ for spin-polarized (in ferromagnetic (FM) and anti-ferromagnetic (AFM) configurations) and non spin-polarized (NSP) calculations.}
\begin{ruledtabular}
\begin{tabular}{ccccc}
System &  &  $a$ (\AA) & $y$ & $z$ \\ 
 \hline
FeSb$_{3}$ & NSP & 9.153 & 0.327  & 0.160 \\
FeSb$_{3}$ &  FM & 9.178& 0.331 & 0.160 \\
FeSb$_{3}$ & AFM & 9.166  & 0.331 & 0.159 \\
FeSb$_{3}$ & exp\cite{Hornbostel} & 9.176 & 0.340 & 0.162 \\
FeSb$_{3}$ & exp $T=10$~K\cite{Mochel} & 9.212 & 0.340 & 0.158 \\
FeSb$_{3}$ & exp $T=300$~K\cite{Mochel} & 9.238 & 0.340 & 0.157 \\
\\
CoSb$_{3}$ & NSP & 9.115 & 0.333  & 0.160 \\
CoSb$_{3}$  & exp\cite{Schmidt} & 9.039 & 0.335 & 0.158 \\
\\
LaFe$_{4}$Sb$_{12}$ & NSP & 9.181 & 0.335 & 0.164 \\
LaFe$_{4}$Sb$_{12}$ & FM & 9.186 & 0.335 & 0.163 \\
 \end{tabular}
\end{ruledtabular}
\end{table}

%%%%%%%%%%%%%%% Results section %%%%%%%%%%%%
\section{Results}\label{sec:results}
\subsection{Structural properties and phase stability}
In Table~\ref{tab:fesb3} we show the evaluated structural parameters for FeSb$_{3}$, CoSb$_{3}$ and LaFe$_{4}$Sb$_{12}$. We find that the non spin-polarized (NSP) calculation for FeSb$_{3}$ yield a smaller lattice constant compared to the result obtained by ferromagnetic (FM) and anti-ferromagnetic (AFM) calculations. The $y$ and $z$ coordinates are however similar in all three cases, especially so for the $z$ coordinate. Compared to experiments there is a very good agreement in the lattice constant between the FM calculation and the experiment by Hornbostel et al.\cite{Hornbostel}, while the lattice constant obtained by M{\"o}chel et al.\cite{Mochel} is much larger compared to our theoretical results. Both of the experimental studies on FeSb$_{3}$ also find $y=0.340$ which is larger than our value of 0.331. However, the $z$ values are very similar. Compared to CoSb$_{3}$ the Fe containing system has a larger lattice constant, while the $y$ and $z$ values are very similar in our calculations. When filling the voids in FeSb$_{3}$ with La, we find that the lattice constant increases by $\sim$0.03~\AA~in the case of NSP calculations and by $\sim$0.01~\AA~in the case of FM calculations, which means that incorporation of La into the lattice is made easier for the FM system.
\par
In addition, we have calculated the energy of the FeSb$_{3}$ phase in relation to FeSb$_{2}$ and elemental Sb according to $\Delta E = E({\rm FeSb}_{3}) - E({\rm FeSb}_{2}) - E({\rm Sb})$, where $E({\rm X})$ is the total energy of system X with X = FeSb$_{3}$, FeSb$_{2}$ and Sb. Here, FeSb$_{2}$ has been evaluated in its orthorhombic ground state with the space group Pnn2, while Sb has been calculated in the A7 trigonal structure with space group R3-mh. For these calculations, we have also increased the density of the k-points mesh to 12$\times$12$\times$12 for FeSb$_{3}$ and FeSb$_{2}$. For Sb the mesh was set to 12$\times$12$\times$6. The resulting very small energy difference of~-0.02~eV/f.u. suggests that the FeSb$_{3}$ phase may be stable or nearly stable towards decomposition into FeSb$_{2}$ and elemental Sb. We note that if SOC is included $\Delta E =-0.01$~eV/f.u. Note that these energies are normalized per formula unit, f.u., of FeSb$_{3}$.
\par

\subsection{Electronic structure}
\begin{figure}[t]
\includegraphics[width=8.5cm]{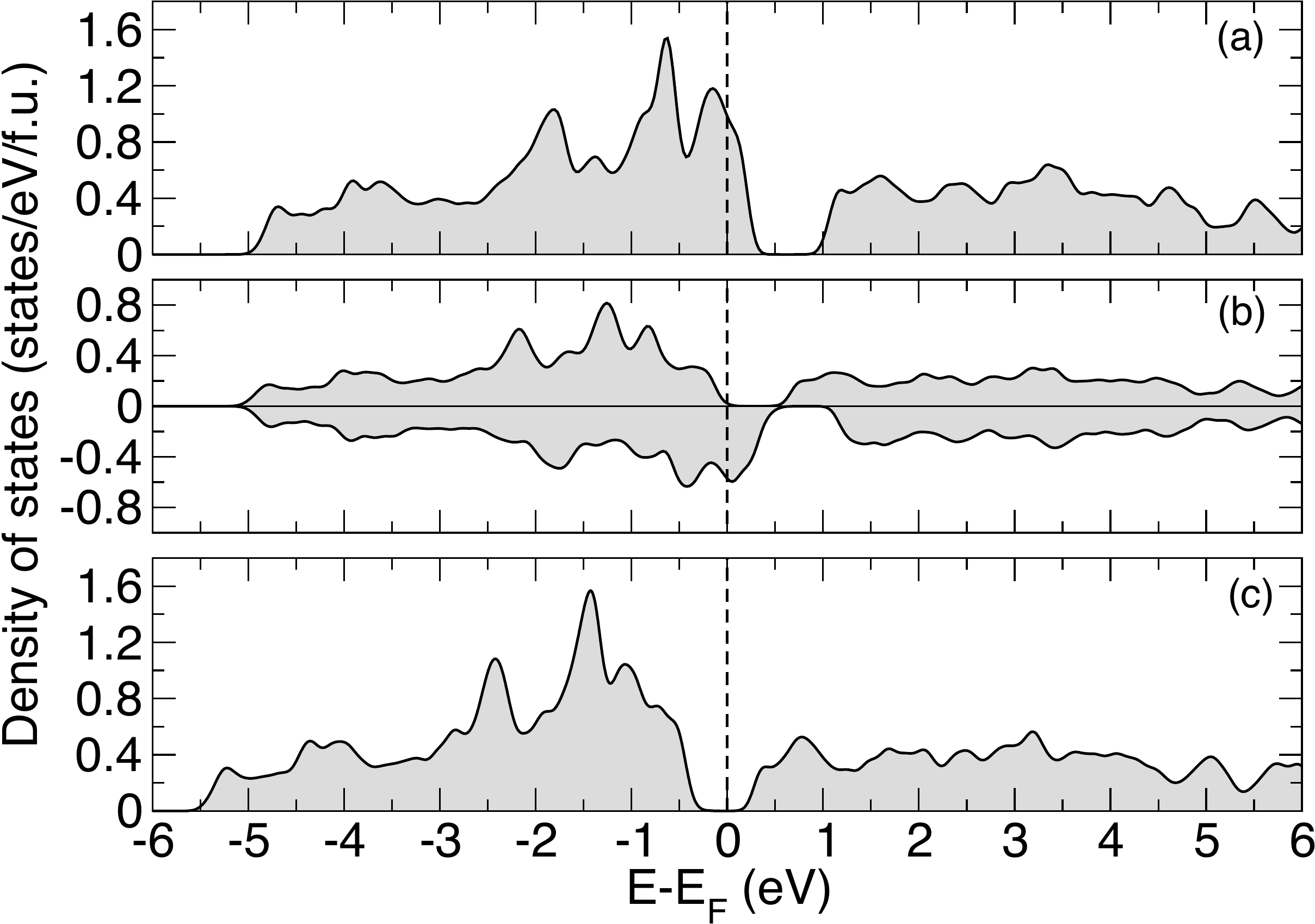}
\caption{\label{fig:electronicDOS} Calculated density of states (DOS) of the upper valence and lower conduction band regions of (a) NSP FeSb$_{3}$, (b) FM FeSb$_{3}$ and (c) CoSb$_{3}$. In (b) the positive DOS are for the spin-up channel and the negative valued DOS are for the spin-down channel. The vertical dashed lines mark the position of the Fermi level.}
\end{figure}
\begin{figure}[t]
\includegraphics[width=8.5cm]{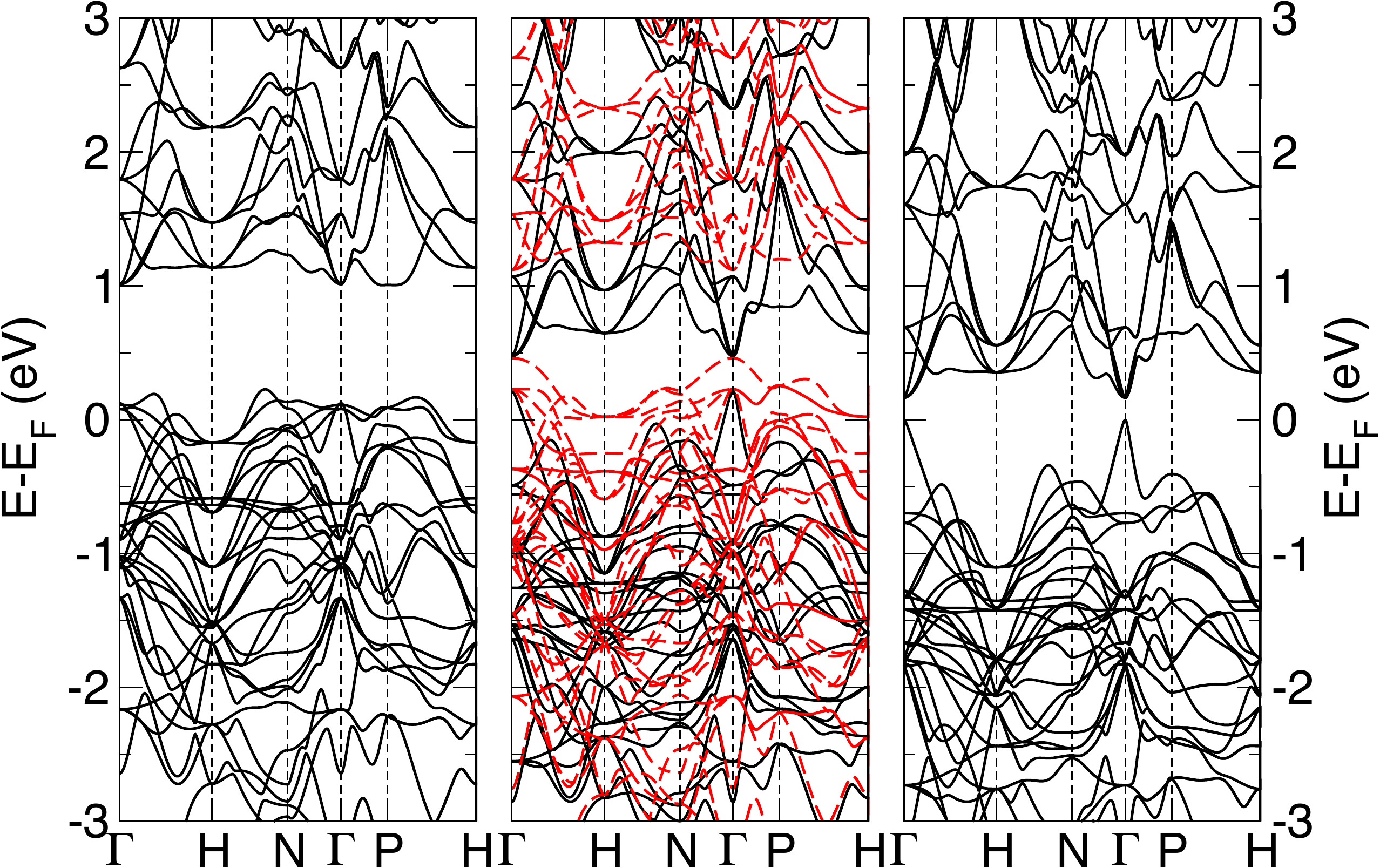}
\caption{\label{fig:electronicBANDS} (Color online) Calculated band structures along high symmetry directions in the Brillouin zone of NSP FeSb$_{3}$ (left panel), FM FeSb$_{3}$ (middle panel) and CoSb$_{3}$ (right panel). In the middle panel the spin-up bands are plotted with solid (black) lines and the spin-down bands are plotted with dashed (red) lines. The zero on the energy scale is the Fermi level,~$E_{F}$.}
\end{figure}
In Fig.~\ref{fig:electronicDOS} we show the electronic density of states (DOS) of NSP FeSb$_{3}$, FM FeSb$_{3}$ and CoSb$_{3}$. It is clear that NSP FeSb$_{3}$ is a metal with a large DOS at the Fermi level, $E_{F}$. The DOS also contains a gap between the valence band edge and higher lying conduction bands. Furthermore, the electronic properties FeSb$_{3}$ is very different from CoSb$_{3}$, which is a semiconductor with a small direct gap at the $\Gamma$ point of about 0.2~eV in agreement with previous theory.\cite{Wee} The large DOS at the Fermi level in the NSP FeSb$_{3}$ makes it energetically favorable for the system to form a ferromagnetic ground state, which can be deduced from the Stoner criterion, i.e. $N(E_{F})I>1$, where $N(E_{F})$ is the DOS at the Fermi level and $I$ is the Stoner exchange parameter. FM FeSb$_{3}$ can be described as a near half-metal, where the spin up channel is almost completely filled with a direct gap at the $\Gamma$ point of about 0.3~eV. The spin down channel has a significant DOS at the Fermi level and up to about 0.5~eV above the Fermi level when a gap of about 0.7~eV is found to the higher lying conduction bands. 
\par
The magnetic moments in the FM and AFM calculations is to a great extent localized on the Fe atoms, with a magnetic moment of $\sim$1.0~$\mu_{B}$/Fe and $\sim$1.1~$\mu_{B}$/Fe respectively. The same is also true for FM LaFe$_{4}$Sb$_{12}$, however, for this system the magnetic moment on the Fe has dropped to $\sim$0.3~$\mu_{B}$/Fe. We also find small induced moments on the Sb atoms (less than 0.01~$\mu_{B}$/atom) and on the La ($\sim$0.05~$\mu_{B}$) in the opposite direction to the moments localized on the Fe atoms. The magnetization energy, i.e. the energy between the FM (or AFM) and the NSP calculations, is for FeSb$_{3}$ -0.28~eV (FM) and -0.13~eV (AFM), which shows that FM FeSb$_{3}$ is the ground state. For this ground state, we obtained a Curie temperature of 175~K which is well below the relevant thermoelectric operating temperatures.  In the case of LaFe$_{4}$Sb$_{12}$ the magnetization energy has been lowered to -0.02~eV which is very small and throughout the remainder of the paper we will only consider NSP LaFe$_{4}$Sb$_{12}$. As expected CoSb$_{3}$ shows no tendency for a magnetic solution. A more detailed account of the magnetic properties of FeSb$_{3}$ and Fe$_{{\rm x}}$Co$_{1-{\rm x}}$Sb$_{3}$ alloys will be presented in Ref.~\onlinecite{magneticFeSb3}.
\par
In addition to the calculated DOS we show in Fig.~\ref{fig:electronicBANDS} the calculated band structures for the NSP FeSb$_{3}$, FM FeSb$_{3}$ and CoSb$_{3}$. It is clear that the electronic structure of FeSb$_{3}$ and CoSb$_{3}$ is significantly different: In CoSb$_{3}$ the highest valence band sticks out above the other bands at the $\Gamma$ point with a quadratic dispersion close to $\Gamma$ that turns into a linear dispersion which extends far out into the Brillouin zone, especially along $\Gamma$ to H.\cite{Smith} This dispersion is also found in FeSb$_{3}$ however, for the NSP FeSb$_{3}$ this band is coexisting with several bands with small dispersion which give rise to the large DOS at the Fermi level. For FM FeSb$_{3}$ the same band is found in the spin up channel, however, here there are several spin down bands at the same energy. 
\par
\begin{table}[t]
\caption{\label{tab:elastic} Calculated elastic constants of NSP, FM and AFM FeSb$_{3}$ in comparison to CoSb$_{3}$ calculated according to Refs.~\onlinecite{elastic1} and \onlinecite{elastic2}. $B = (1/3)(c_{11}+2c_{12})$ is the bulk modulus and $c' = (1/2)(c_{11}-c_{12})$ is the tetragonal shear constant.}
\begin{ruledtabular}
\begin{tabular}{ccccccc}%ccccc}
System &  $c_{11}$ (GPa) & $c_{12}$ (GPa) & $c_{44}$ (GPa) & $B$ (GPa) & $c'$ (GPa) \\
 \hline
 NSP &  166.0 & 41.7 & 22.0 & 83.1 & 62.2 \\
  FM &  166.0 & 37.1 & 35.0 & 80.1 & 64.5 \\
 AFM  & 156.6 & 37.0 & 31.6 & 76.9 & 59.8 \\
CoSb$_{3}$  & 181.8 & 36.5 & 49.4 & 84.9 & 72.7\\
 \end{tabular}
\end{ruledtabular}
\end{table}
\subsection{Elastic constants and the velocity of sound}
In Table~\ref{tab:elastic} we show the calculated elastic constants of FeSb$_{3}$ and CoSb$_{3}$. We find that essentially all elastic constants is smaller in FeSb$_{3}$ compared to CoSb$_{3}$. It is only $c_{12}$ that is larger in FeSb$_{3}$, although the differences in the elastic constants in the two systems are small. The bulk modulus $B$ and the tetragonal shear constant $c'$ are both larger in CoSb$_{3}$ compared to FeSb$_{3}$. We can therefore conclude that the FeSb$_{3}$ framework is softer than CoSb$_{3}$, which is in agreement with experimental findings\cite{Mochel} as well as with our analysis of the lattice dynamics, see below. M{\"o}chel et al.\cite{Mochel} have reported the bulk modulus of FeSb$_{3}$ to be 47.9(1)~GPa\cite{Mochel} which is much smaller than the values obtained by us here. This could be related to the difference in lattice constant between our calculation and the experiment, see Table~\ref{tab:fesb3}, where the lattice constant of Ref.~\onlinecite{Mochel} is larger than the theoretical value. Since the bulk modulus can be expressed as $B=(1/V)\partial^2 E/\partial V^2$, a large difference in the volume will affect the bulk modulus. For CoSb$_{3}$ the experimental bulk modulus is 83.2(1)~GPa\cite{Mochel} which is in excellent agreement with our result. 
\par
The difference in the size of the elastic constants  will affect the velocity of sound in the materials, since the velocity of sound in a given direction is given by $c=\sqrt{c_{\rm eff}/\rho}$, where $c_{\rm eff}$ is a linear combination of elastic constants depending on the direction and $\rho$ is the mass density of the material. A simple analysis yields that the increased stiffness of CoSb$_{3}$ compared to FeSb$_{3}$ means that sound waves travel more readily in CoSb$_{3}$. Since the velocity of sound is in turn related to the thermal conductivity of the material we may anticipate that the thermal conductivity is lower in FeSb$_{3}$ compared to CoSb$_{3}$. However, this has to be supported in a more rigorous fashion. 

\begin{figure}[t]
\includegraphics[width=8cm]{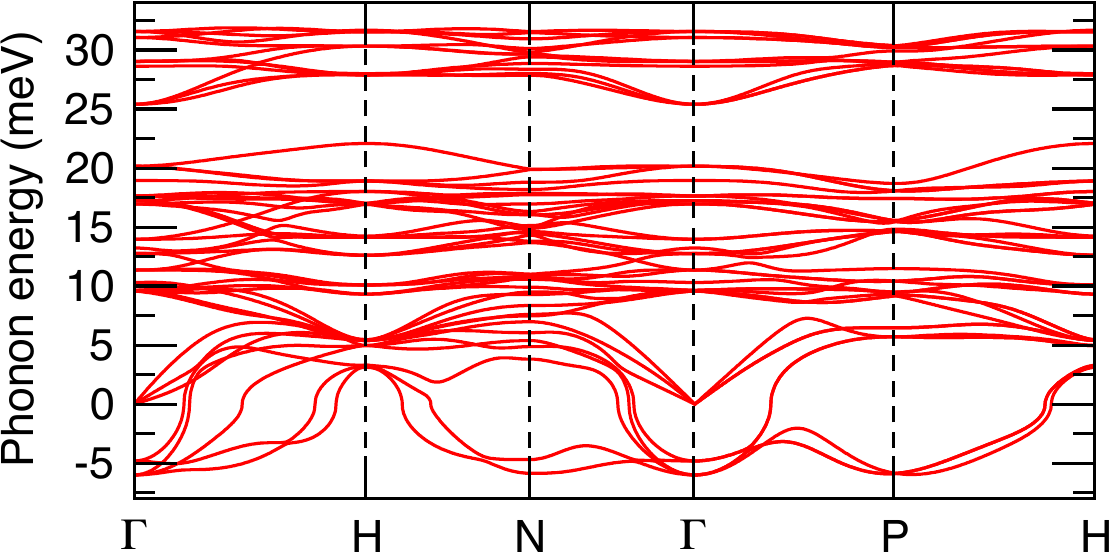}
\caption{\label{fig:imaginary} (Color online) The resulting phonon dispersions along high symmetry directions in the Brillouin zone based on a non spin-polarized density functional calculation of FeSb$_{3}$. Imaginary phonon energies are plotted with negative values. Note the imaginary phonon energies along all directions shown.}
\end{figure}

\begin{figure}[t]
\includegraphics[width=8cm]{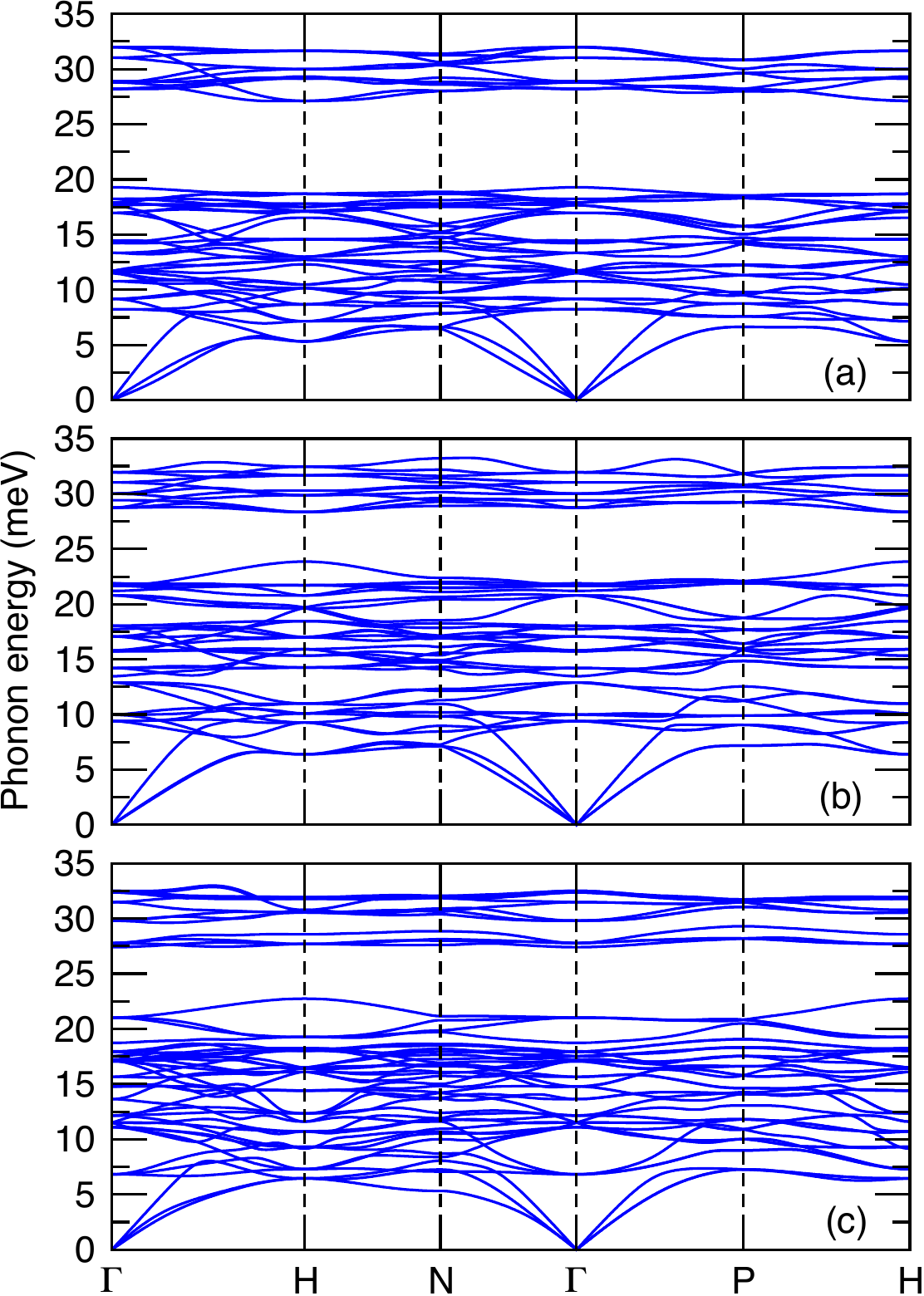}
\caption{\label{fig:bands} (Color online) Calculated phonon dispersions for (a) FM FeSb$_{3}$, (b) CoSb$_{3}$ and  (c) LaFe$_{4}$Sb$_{12}$ along high symmetry directions in the Brillouin zone.}
\end{figure}

\begin{figure}[t]
\includegraphics[width=8cm]{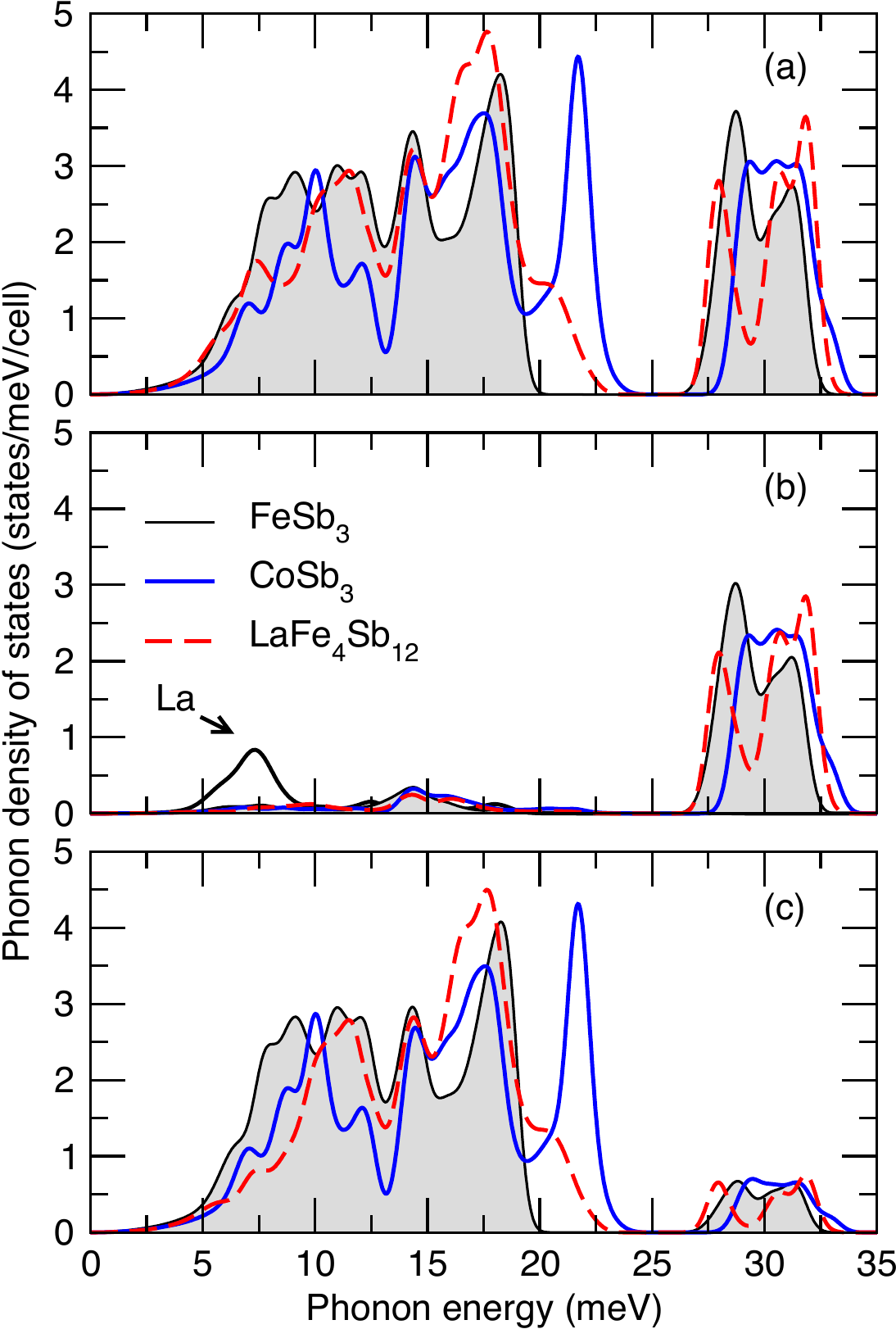}
\caption{\label{fig:phononDOS} (Color online) Phonon density of states (PDOS) of FeSb$_{3}$, CoSb$_{3}$ and LaFe$_{4}$Sb$_{12}$, (a) total PDOS, (b) projected PDOS on metallic species, i.e. Fe, Co and La, (c) projected PDOS on Sb. The projected PDOS onto La is given by the thick black line in (b). For clarity, the PDOS on atoms in the FeSb$_{3}$ system are shaded in grey. Note that the unit is states/meV/cell, where the cell is the either Fe$_{4}$Sb$_{12}$, Co$_{4}$Sb$_{12}$ or LaFe$_{4}$Sb$_{12}$.}
\end{figure}
\par
\subsection{Lattice dynamics}
In Fig.~\ref{fig:imaginary} we show the calculated phonon dispersion energies of FeSb$_{3}$ obtained from a non spin-polarized (NSP) density functional calculation. Several phonon modes have imaginary phonon energies and it is clear that NSP FeSb$_{3}$ is dynamically unstable within the harmonic approximation. By allowing for a FM solution this behavior will change and all phonon modes will have real values. This is shown in Fig.~\ref{fig:bands}, where we present the calculated phonon energies of FM FeSb$_{3}$ and compare the results with the phonon energy dispersions for CoSb$_{3}$ and LaFe$_{4}$Sb$_{12}$. For the remainder of the paper we will focus only on FM FeSb$_{3}$ and the FM descriptor will be omitted throughout.
\par
In Fig.~\ref{fig:bands}, it is clear that the phonon modes can be divided into two regions in terms of energy: The lower region which contains both acoustic and optical modes, which in the case of FeSb$_{3}$ lies in the energy range of 0~to~20~meV, and the upper optical region, which for FeSb$_{3}$ lies in between 27~to~33~meV. In CoSb$_{3}$ the lower region reaches $\sim$25~meV which is about 5~meV higher in energy compared to FeSb$_{3}$. This suggests that FeSb$_{3}$ is softer than CoSb$_{3}$ which agrees well with recent experimental findings\cite{Mochel}. In addition, we find that the upper optical region in LaFe$_{4}$Sb$_{12}$ is split in two regions with a gap at 27~meV, which is dramatically different from the behavior in the binary systems where this region contains modes with dispersions that cross and overlap each other. For CoSb$_{3}$ there is a narrow gap at about 13~meV which splits the lower phonon energy region. Among the systems studied here, this gap is unique to CoSb$_{3}$. The existence of this gap has been observed previously by both theory and experiment. \cite{Feldman,Feldman2,Feldman3,Hermann,Kendziora} We also note a clear signature of avoided crossing\cite{Toberer} of the acoustic and low lying optical modes in LaFe$_{4}$Sb$_{12}$ along the  line connecting $\Gamma$ to P which does not exist in the binary systems. The acoustic phonon modes are also found to be non degenerate along most directions in the Brillouin zone. It is only along $\Gamma$ to P where the two lowest energy phonon modes are degenerate for all three systems in our study.
\par
In Fig.~\ref{fig:phononDOS} we have plotted the total and projected phonon density of states (PDOS) for FeSb$_{3}$, CoSb$_{3}$ and LaFe$_{4}$Sb$_{12}$. In FeSb$_{3}$ the lower energy region between 0 and 20~meV is dominated by Sb and the upper region between 27 to 33~meV is dominated by Fe. We find that the PDOS of Fe has two small broad peaks at $\sim$7~meV and  $\sim$15~meV, as well as a larger feature between 27 and 33~meV which has two clear peaks, see panel (b) in Fig.~\ref{fig:phononDOS}. The Sb projected PDOS is more or less flat between 0 and 20~meV with two valleys at $\sim$13~meV and $\sim$16~meV. These observations are in good agreement with available experimental PDOS measured by M{\"o}chel et al.\cite{Mochel} Compared to CoSb$_{3}$, also plotted in Fig~\ref{fig:phononDOS}, the PDOS of FeSb$_{3}$ show the same behavior as the phonon dispersions that the lower phonon energy region is narrower in FeSb$_{3}$. In addition, in CoSb$_{3}$ the Sb projected PDOS shows distinct features, such as two deep valleys at $\sim$13~meV and $\sim$20~meV as well as a very distinct peak at $\sim$22~meV. The very deep valley at 13~meV coincides with the gap in the phonon dispersions for CoSb$_{3}$ shown in Fig.~\ref{fig:bands}. The peak at 22~meV and valley at 20~meV mark significant differences to the Sb PDOS in FeSb$_{3}$. For the PDOS projected onto Co we find that it also show smaller features at the same energies as the Fe PDOS in FeSb$_{3}$, however the feature at large phonon energies is shifted to slightly larger energies for the Co PDOS compared to the case of Fe in FeSb$_{3}$, and this feature is flat which is a distinct difference to Fe in FeSb$_{3}$. The two binary skutterudite systems therefore show distinct differences in their lattice dynamics and especially so for the dynamics of the Sb atoms in the two systems.
\par
When La is inserted into FeSb$_{3}$, we find that the total PDOS from 0 up to about 7.5~meV are very similar, however, when projected onto Sb there is a marked difference between the filled and unfilled systems. The Sb PDOS has been reduced in response to the filling in this region. This reduction in the Sb PDOS for LaFe$_{4}$Sb$_{12}$ coincides with the appearance of a filler mode due to vibrations of the La centered about 7~meV, see Fig.~\ref{fig:phononDOS}. We also find features in the Sb PDOS for the filled system at this energy which suggests that the La and Sb vibrations are connected and not independent which is in agreement with experimental observations\cite{Koza}. In addition to the large La peak at about 7~meV, we find several small features in the La PDOS at about 12.5~meV and 18~meV in agreement with previous theory and experiment.\cite{Feldman} Furthermore, we find that the PDOS of the filled system is broadened compared to the unfilled FeSb$_{3}$ and that the Fe PDOS region at high phonon energies are split with a deep valley at just below 30~meV. This latter feature is due to the gap in the high energy optical modes which is clearly visible in Fig.~\ref{fig:bands}.

%%%%%%%%%%%%% Conclusions %%%%%%%%%%%%%%%
\section{Summary and Conclusions}\label{sec:conclusions}
We have performed density functional calculations on the FeSb$_{3}$ skutterudite system and analyzed its electronic structure and lattice dynamics. The electronic structure has been compared in detail with CoSb$_{3}$ while for the lattice dynamics comparison has also been done with the filled LaFe$_{4}$Sb$_{12}$ skutterudite phase. We find that FeSb$_{3}$ is a near semi-metal with a ferromagnetic ground state and a Curie temperature of 175~K. The non spin-polarized FeSb$_{3}$ is a metal with a large DOS at the Fermi level which facilitates the ferromagnetic ground state.
\par
We find that in order to obtain a dynamically stable system it is required to evaluate the lattice dynamics for the ferromagnetic ground state of FeSb$_{3}$. Phonon calculations for the non spin-polarized phase yield unstable phonon modes along most high symmetry directions. We note that the calculations of the lattice dynamics have been done within the harmonic approximation at $T=0$~K. However, the harmonic approximation is known to fail for systems that are stabilized at finite temperatures, since anharmonic effects are not included in the calculation. Famous examples are the high temperature phases of Zr,\cite{Hellman,Petros} Ti\cite{Petros} and Hf\cite{Petros} which all have the bcc structure at elevated temperatures even though the bcc structure for these systems is dynamically unstable in the harmonic approximation. The low T$_{c}$ for FeSb$_{3}$ combined with the fact that the phase is made experimentally at $\sim400$~K\cite{Hornbostel,Mochel} suggest that the non spin-polarized phase will be stabilized at finite temperature in a similar fashion as the examples mentioned above.
\par
Compared to CoSb$_{3}$ we find that the lattice dynamics in FeSb$_{3}$ differs in several respects, for example: The Sb PDOS in FeSb$_{3}$ is more flat and narrower compared to CoSb$_{3}$ which shows more features in the PDOS, especially a significant peak about 22~meV that is not present in FeSb$_{3}$. In addition, the PDOS of Fe at large phonon energies show a clear two peak feature which is not the case in CoSb$_{3}$ which is more flat. We also find that FeSb$_{3}$ is softer than CoSb$_{3}$ based on the form of the phonon energy dispersions and PDOS as well as from calculations of the elastic constants of the two systems.
\par
When compared to filled LaFe$_{4}$Sb$_{12}$, we find a hardening at low phonon energies for the filled system of the Sb PDOS which coincides with the La PDOS peak centered at 7~meV. This hardening of the Sb PDOS is in perfect agreement with the experimental result of M{\"o}chel et al.\cite{Mochel}
\par
Furthermore, we find that FeSb$_{3}$ is mechanically stable, i.e. have $B$, $c'$ and $c_{44}$ greater than zero, as well as dynamically stable in its bulk form. This suggests that it should be in principle possible for a process route to be established in order to produce bulk samples of binary FeSb$_{3}$.

%%%%%%% 
\section{Acknowledgements}
This work was financed through the EU project NexTec, VR (the Swedish Research Council), and SSF (Swedish Foundation for Strategic Research). The computations were performed on resources provided by the Swedish National Infrastructure for Computing (SNIC) at the National Supercomputer Centre in Link{\"o}ping (NSC).

\end{document}